\documentclass[final,12pt]{elsarticle}
\usepackage{graphicx}
\usepackage{epsfig}
\usepackage{amssymb}
\usepackage{amsthm}
\usepackage{latexsym}
\usepackage{rotating}
\usepackage{longtable}

\journal{Journal of Molecular Spectrosoopy}

\begin{document}

\begin{frontmatter}

\title{High precision measurement of ultraweak transitions of the
a${^1}{\triangle}_g{\leftarrow}$X $^3{\Sigma}^{-}_g$ $(0,0)$ band
of molecular oxygen}

\author[label11]{Dipankar Das}

\address[label11]{Department of Physics, University of Otago, Dunedin, New Zealand}

\begin{abstract}
We have used a highly sensitive cavity-enhanced frequency
modulation spectroscopy technique to measure different parameters
of the ultraweak transitions of a${^1}{\triangle}_g{\leftarrow}$X
$^3{\Sigma}^{-}_g$ $(0,0)$ band of molecular oxygen in the range
7640 cm$^{-1}$ to 7917 cm$^{-1}$. The self-broadened half-width
and air-broadened half-width of the transitions have been measured
for three different pressures for both $^{16}$O$_2$ and
$^{18}$O$_2$. To measure the line intensity and self-broadened
half-width we have used ultra pure oxygen sample and air-broadened
half-width was measured with dry air sample. The $(0,0)$ band of
$^{16}$O$_2$ and $^{18}$O$_2$ show weak quadrupole transitions
with line intensities ranging from $1\times10^{-30}$ to
$1.9\times10^{-28}$ cm/molecule. The measurements are in excellent
agreement with the recent measurement of Rothman et. al.
\cite{OSI10}.
\end{abstract}


\end{frontmatter}

\section{Introduction}
The precise measurement of different parameters in the infrared
spectrum in the a${^1}{\triangle}_g{\leftarrow}$X
$^3{\Sigma}^{-}_g$ $(0,0)$ band of oxygen is of fundamental
interest to the atmospheric scientists for modelling and
calculating different physical parameters like radiation transfer,
heat distribution due to the greenhouse gases, ozone abundances
and hence providing a comparative study of different atmospheric
models available. Oxygen transitions are often used as a
calibrator on the spectra obtained from the satellite instruments.
Furthermore taking the ratio of column abundances of CO$_2$ or
CH$_4$to that of O$_2$ cancels many common systematic errors which
are related to the instruments \cite{WTB06}. So there is a
constant demand that all the spectroscopic parameters
corresponding to this a-X $(0,0)$ transition are precisely
measured.

Due to the importance, magnetic dipole transitions in this band
were tabulate in the HITRAN database \cite{RGB09}. Since these
parameters have been continuously updated through different
experimental results as HITRAN was evolving. The intensities of
the a${^1}{\triangle}_g{\leftarrow}$X $^3{\Sigma}^{-}_g$ $(0,0)$
band for $^{16}$O$_2$ in HITRAN (modified for HITRAN2000 using
experimental parameters from Lafferty et al. \cite{LSL98}).
However it were found to be inadequate for atmospheric retrievals
such as the one in Ref.\cite{WTB06}. The updated list of
$^{16}$O$_2$ intensities recently adopted for HITRAN was derived
by Orr-Ewing based on experimental results from Newman et al.
\cite{NLO99}. These intensities are much accurate in application
to atmospheric retrievals \cite{WTB06}. In order to improvise the
retrievals, Washenfelder et al. \cite{WTB06} used an empirical
scaling to HITRAN2008 intensities of $^{16}$O$^{18}$O. These
intensity improvements have served as a basis for an interim
update of the HITRAN oxygen file that was placed on the HITRAN
website in November 2009. There is still space for substantial
improvement of spectral parameters of the
${^1}{\triangle}_g{\leftarrow}$X $^3{\Sigma}^{-}_g$ $(0,0)$ band
of molecular oxygen. Recently Gordon et al. \cite{GKC10} have
shown the importance of including quadrupole lines in this region
to the HITRAN line list to improve the residuals of the solar
absorption spectrum.

The line position of  the a${^1}{\triangle}_g{\leftarrow}$X
$^3{\Sigma}^{-}_g$ $(0,0)$ band transitions of $^{16}$O$_2$ in
HITRAN, the X $^3{\Sigma}^{-}_g$ spectroscopic constant from
Rouilli$\acute{e}$ et al. \cite{RMS92} and a${^1}{\triangle}_g$
rotational constants determined from the microwave spectrum by
Hilling et al. \cite{HCR85} were used. Gamache et al. \cite{GGR98}
describe that the term energy of a${^1}{\triangle}_g$ was taken
from Krupenie \cite{KRU72}. The term values of Krupenie were
corrected in HITRAN to match the unpublished FTS spectrum measured
by Brault \cite{BRU82}. In HITRAN2008, line position of
$^{16}$O$^{18}$O were calculated using ground state constant of
Yamomoto et al. \cite{MYA91} and excited -state rotational
constants reported by Herzberg et al. \cite{HHE47}.

Though the HITRAN line positions are well accepted universally
till today, a much more precise measurement of the
a${^1}{\triangle}_g{\leftarrow}$X $^3{\Sigma}^{-}_g$ $(0,0)$ band
is required. Furthermore, a recent measurement of Rothman et al.
\cite{OSI10} and the measurement Cheah et al. \cite{CLO00} differs
significantly though Chech et al. \cite{CLO00} claim that their
measurement are better than those in HITRAN. This discrepancies
also invite attention for further investigation and precise
measurement parameters of this a${^1}{\triangle}_g{\leftarrow}$X
$^3{\Sigma}^{-}_g$ $(0,0)$ band transitions.

In this paper, we present measurement of the
a${^1}{\triangle}_g{\leftarrow}$X $^3{\Sigma}^{-}_g$ $(0,0)$ band
electronic transitions using a high sensitive cavity-enhanced
frequency modulation spectroscopy. The present work measures
different parameters i.e. the self-broadened half-width,
air-broadened half-width, line-center frequency and line intensity
of the O$_2$ a-X $(0,0)$ band using ultra pure oxygen sample and
dry air sample at three different pressures (200 mbar, 400 mbar
and 600 mabr). Our measurement of the line positions of the
transitions are used to determine improved molecular constants
which are of significance for the correct interpretation of the
atmospheric phenomena.

\section{Experimental setup}
To record our data we use the ultrasensitive technique of
noise-immune cavity-enhanced optical heterodyne molecular
spectroscopy (NICE-OHMS) \cite{JLJ98,JLD99}. NICE-OHMS is a
technique in which frequency modulation spectroscopy (FMS) is
performed within a high-finesse cavity, thereby increasing the
effective sample path length by $\frac{2\times finesse}{\pi}$. The
laser frequency is locked to a longitudinal cavity mode and the
modulation frequency is locked to exactly match the cavity free
spectral range (f$_{FSR}$). This ensures all the frequency
components of the light are affected identically by the cavity.
The light transmitted by the cavity is incident on a fast
photodiode and the output is demodulated to produce a
cavity-enhanced FMS signal. A second level of modulation, at low
frequency, is applied to the cavity (and hence the laser), and
this signal is demodulated by a lock-in amplifier to produce the
NICE-OHMS signal.

Details of our NICE-OHMS experimental setup are described in an
earlier work from our laboratory \cite{NLW04}. Briefly, our
experimental setup uses an external cavity diode laser with a
wavelength range from 1260 nm to 1310 nm, and a high finesse
cavity with a free spectral range of 540 MHz. The cavity has a
finesse $\simeq$ 16000 and thus an effective path length of over
2.8 km. The cavity output mirror is mounted on a piezoelectric
transducer (PZT) to allow scanning of the cavity length. The
cavity lies within an ultrahigh vacuum chamber, evacuated to a
base pressure of 10$^{-7}$ mbar before being filled with molecular
oxygen sample (ultra pure oxygen) or dry air sample. The
intra-cavity pressure was measured using a capacitance manometer
(Pfeiffer Vacuum TPG 261) with a range of 1-1100 mbar and a
precision of 0.1$\%$. Over the period our measurements were taken
the temperature of the vacuum chamber drifted with the variation
in room temperature between 22.4 and 23.3$^{\circ}$C.

We used a National Instruments PCI-6259 M-series data-acquisition
card to perform frequency scans and to collect our NICE-OHMS data.
A voltage ramp from an analog output was fed to the cavity PZT,
producing a scan of the cavity length and hence the laser
frequency. Each scan took 40s and spanned up to 0.42 cm$^{-1}$.
The cavity PZT does not produce a linear scan with voltage, so a
frequency scaling measurement was made for each data set. Each 42
cm$^{-1}$ wide spectrum recorded with one single laser (New Focus
TLB 6324-D Velocity Laser) was calibrated independently on the
basis of the wavelength values provided by a Michelson-type
wavemeter (Bristol 621A NIR, absolute accuracy $\pm$0.002
cm$^{-1}$ @ 10,000 cm$^{-1}$). Furthermore we applied lambdameter
reading corrections as described in a earlier work by Perevalov
\cite{PJR06}. After the correction, the entire spectrum become
linear in terms of frequency. For the data presented in this
paper, the output of the lock-in amplifier (the NICE-OHMS signal)
was averaged over 12 consecutive scans. The typical uncertainty in
the line position is estimated to be less than 1$\times$10$^{-3}$
cm$^{-1}$.

We used Matlab to perform a fit of the collected data to a
theoretical NICE-OHMS lineshape function \cite{NLW04}. The
theoretical lineshape is based on a Voigt lineshape and accounts
for the effect of the two levels of frequency modulation. As the
experiment is performed at room temperature, we calculate the
Gaussian width at 23$^{\circ}$C  which causes the Doppler
broadening, and the fitting routine produced best fit parameters
for the peak absorption, the Lorentzian half-width due to pressure
broadening, and thus the line intensity. We estimate the
experimental precision for each of the transitions on a case by
case basis, determining the range over which fitted parameters are
reasonable. A detailed discussion of the sources of error is
discussed in a earlier work from our laboratory \cite{NLW04}. For
stronger, well-resolved, transitions we estimate an experimental
precision of 4$\%$ for the linewidth and 7$\%$ for the line
intensity, whereas for the weakest, least resolved, transitions we
have estimated errors of up to 40$\%$ for the linewidth, and
70$\%$ for the line intensity. In addition, the uncertainty in our
detection system gain in 8$\%$ so the overall accuracy of the line
intensities is there experimental precision plus 8$\%$, while the
overall accuracy of the linewidths is equal to the experimental
precision. For our experiment we have used ultra pure oxygen and a
dry air mixture to measure the self-broadened and air-broadened
half-width respectively.

\section{Result and analysis}
\begin{figure}
\centering \scalebox{0.7}{\includegraphics{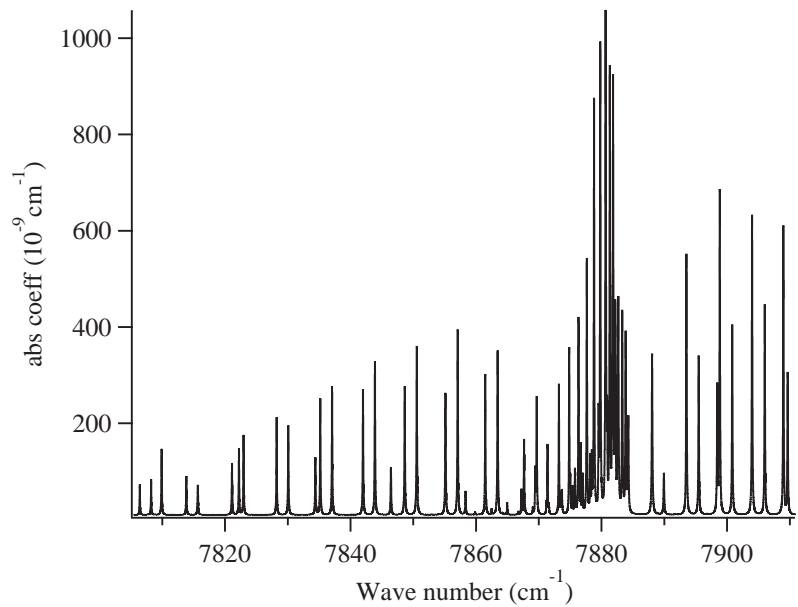}}
\caption{\label{o2transition} Overview of the transitions in the
a${^1}{\triangle}_g{\leftarrow}$X $^3{\Sigma}^{-}_g$ band of
oxygen recorded by CW-CRDS (P = 400 mbar, T= 300 K).}
\end{figure}

The overview of the transitions in the
a${^1}{\triangle}_g{\leftarrow}$X $^3{\Sigma}^{-}_g$ band of
oxygen is shown in Fig.~\ref{o2transition}. The transitions
reported within this paper lie between 7640 to 7917 cm$^{-1}$ and
it is clear that from the overview spectra that the line strength
of these transitions are very low. We were able to measure only
the symmetric isotopologues i.e. $^{16}$O$_2$ and $^{18}$O$_2$,
since the line strengths for the transitions of asymmetric
isotopologues are much lower than the resolution limit of our
experimental set up.

\subsection{Lines of $^{16}$O$_2$}

\begin{figure}
\centering
\begin{minipage}{.65\textwidth}
  \centering
  \includegraphics[width=.9\linewidth]{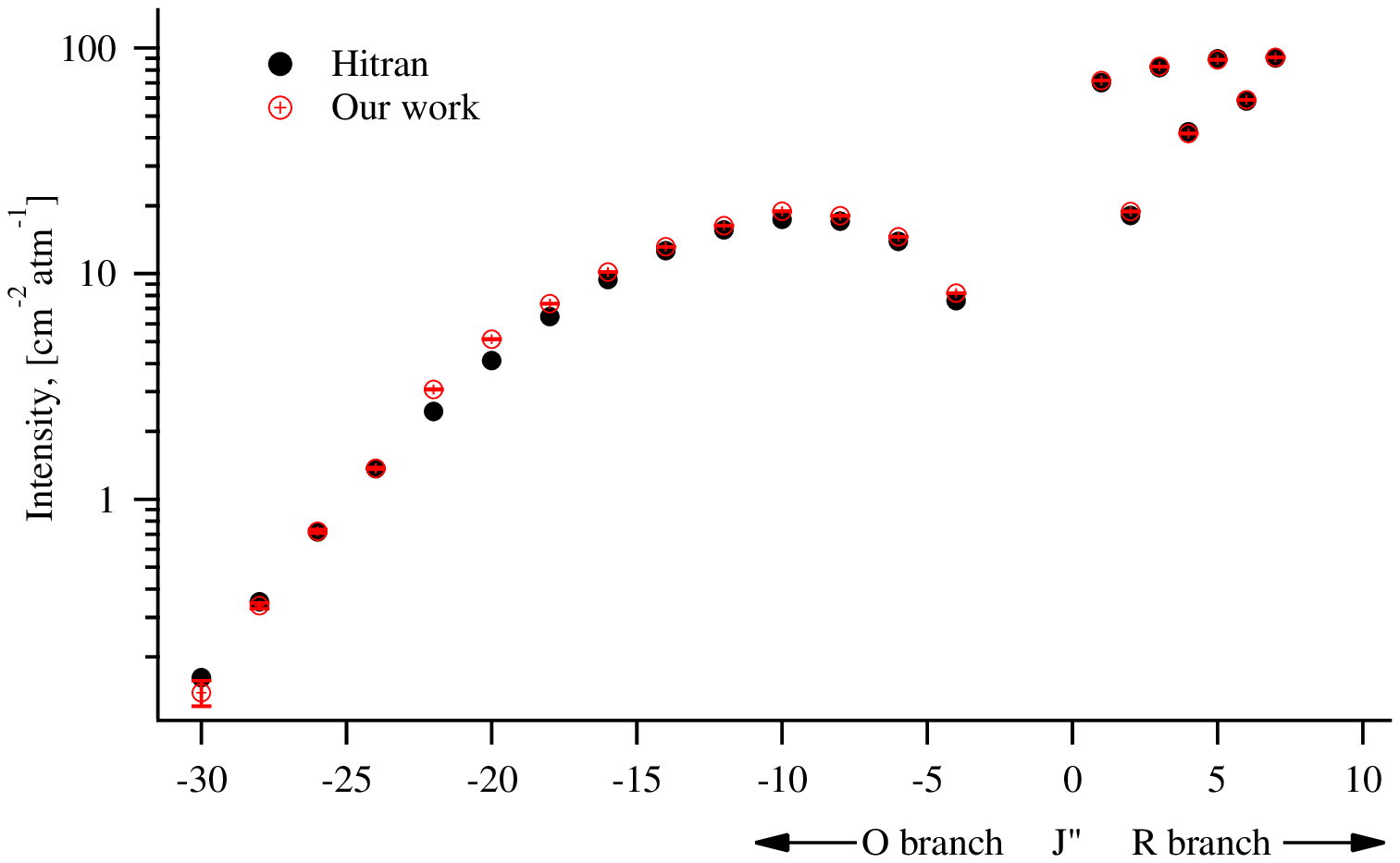}
  \caption{\label{o16O-Rbranch} Intensity of the $O$, $R$ branch of $^{16}$O$_2$.}
\end{minipage}%
\begin{minipage}{.65\textwidth}
  \centering
  \includegraphics[width=.9\linewidth]{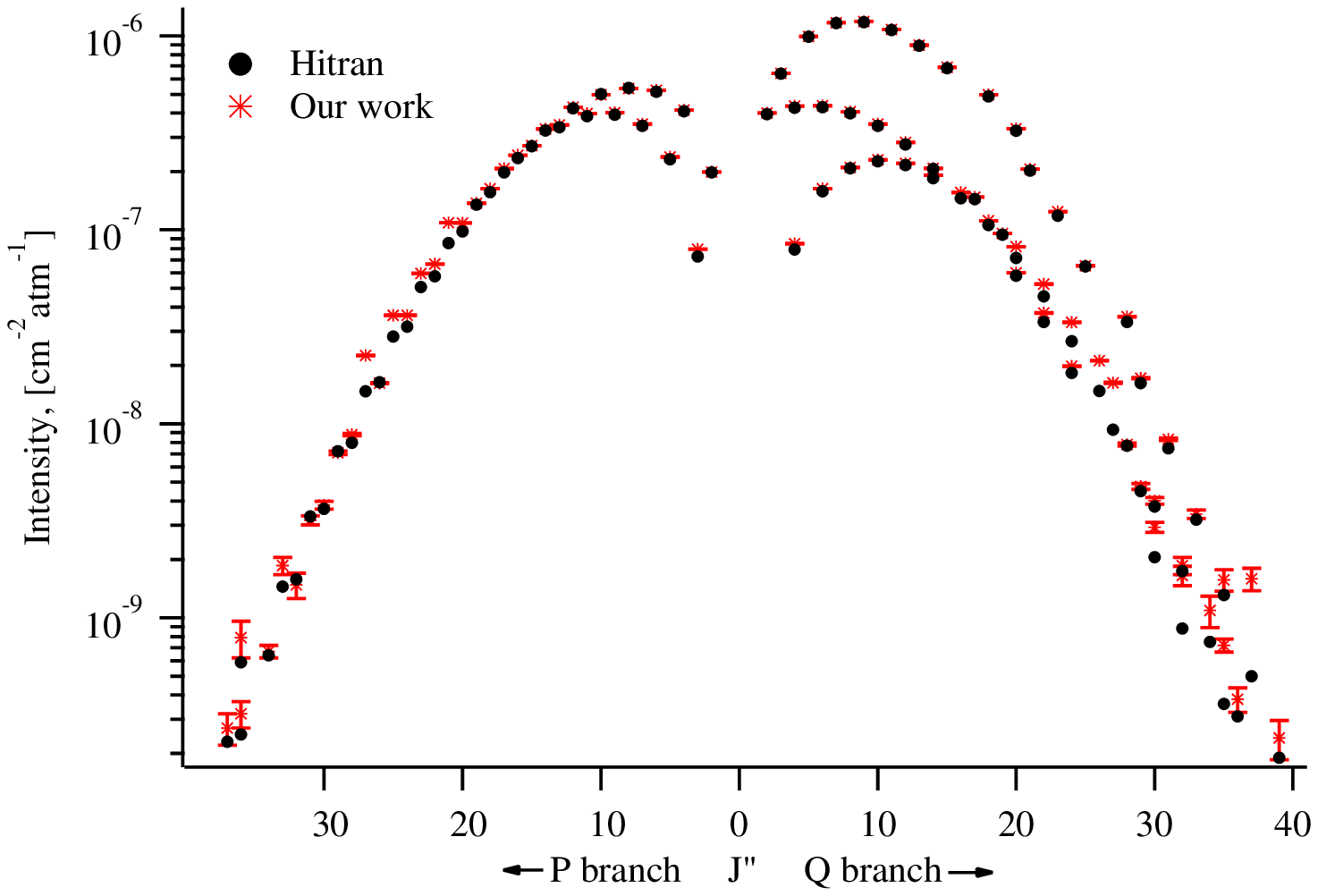}
  \caption{\label{o16P-Qbranch} Intensity of the $P$, $Q$ branch of $^{16}$O$_2$.}
\end{minipage}
\end{figure}

\begin{figure}
\centering
\begin{minipage}{.65\textwidth}
  \centering
  \includegraphics[width=.8\linewidth]{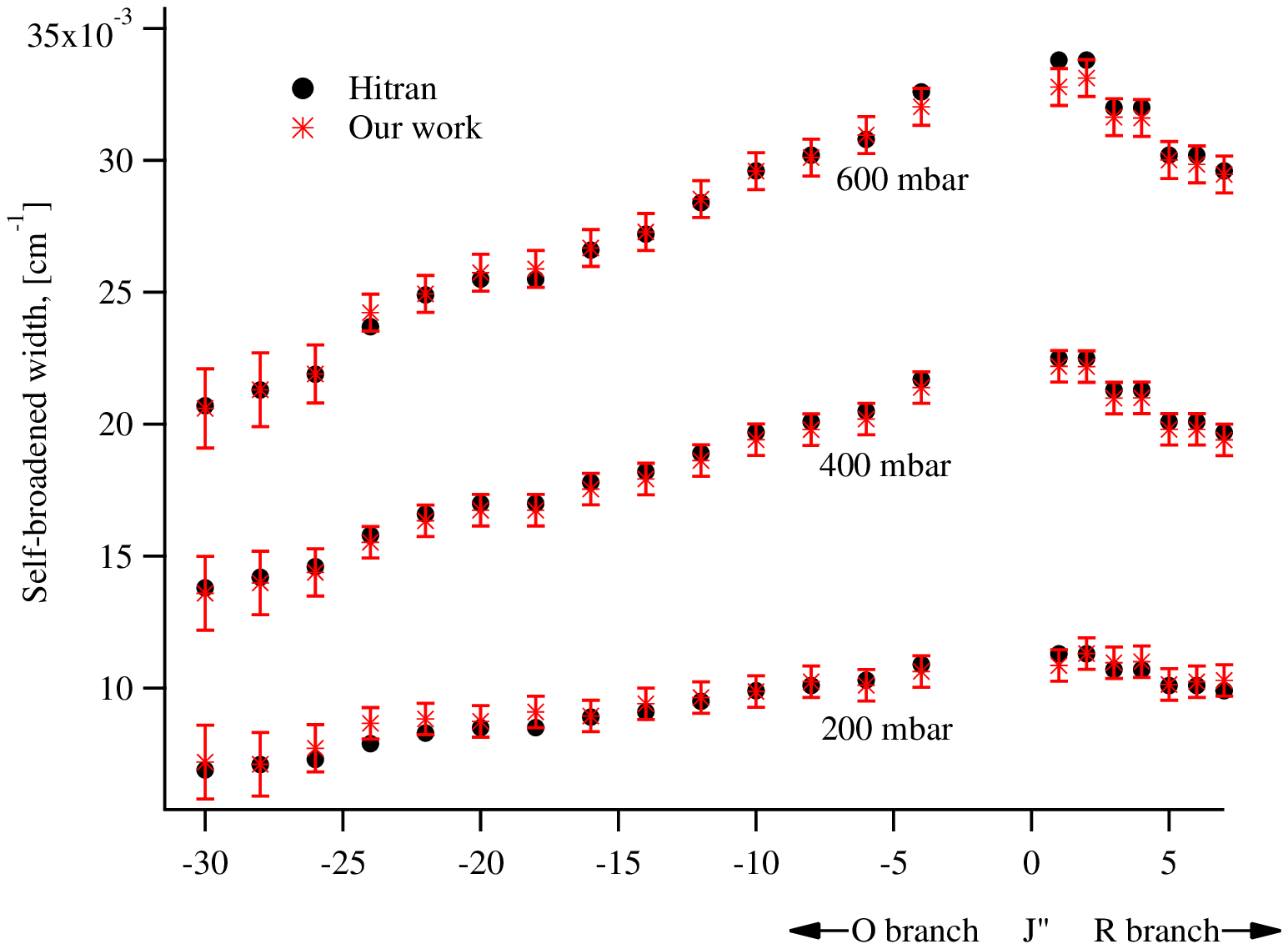}
  \caption{\label{o16ORselfbroadened} Width of $O$, $R$ branch $^{16}$O$_2$.}
\end{minipage}%
\begin{minipage}{.65\textwidth}
  \centering
  \includegraphics[width=.8\linewidth]{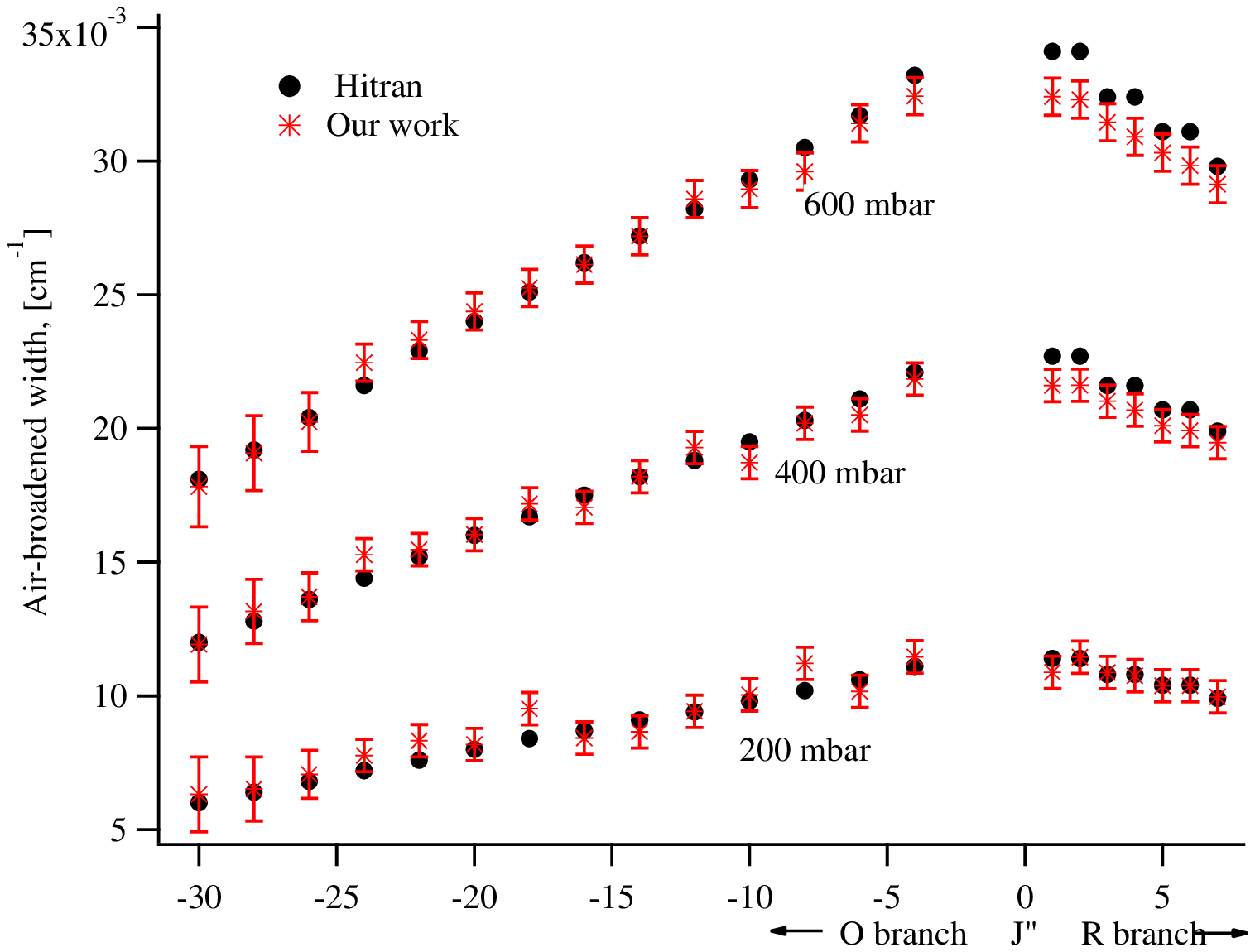}
  \caption{\label{o16ORairbroadened} Width of $O$, $R$ branch $^{16}$O$_2$.}
\end{minipage}
\end{figure}

\begin{figure}
\centering
\begin{minipage}{.65\textwidth}
  \centering
  \includegraphics[width=.8\linewidth]{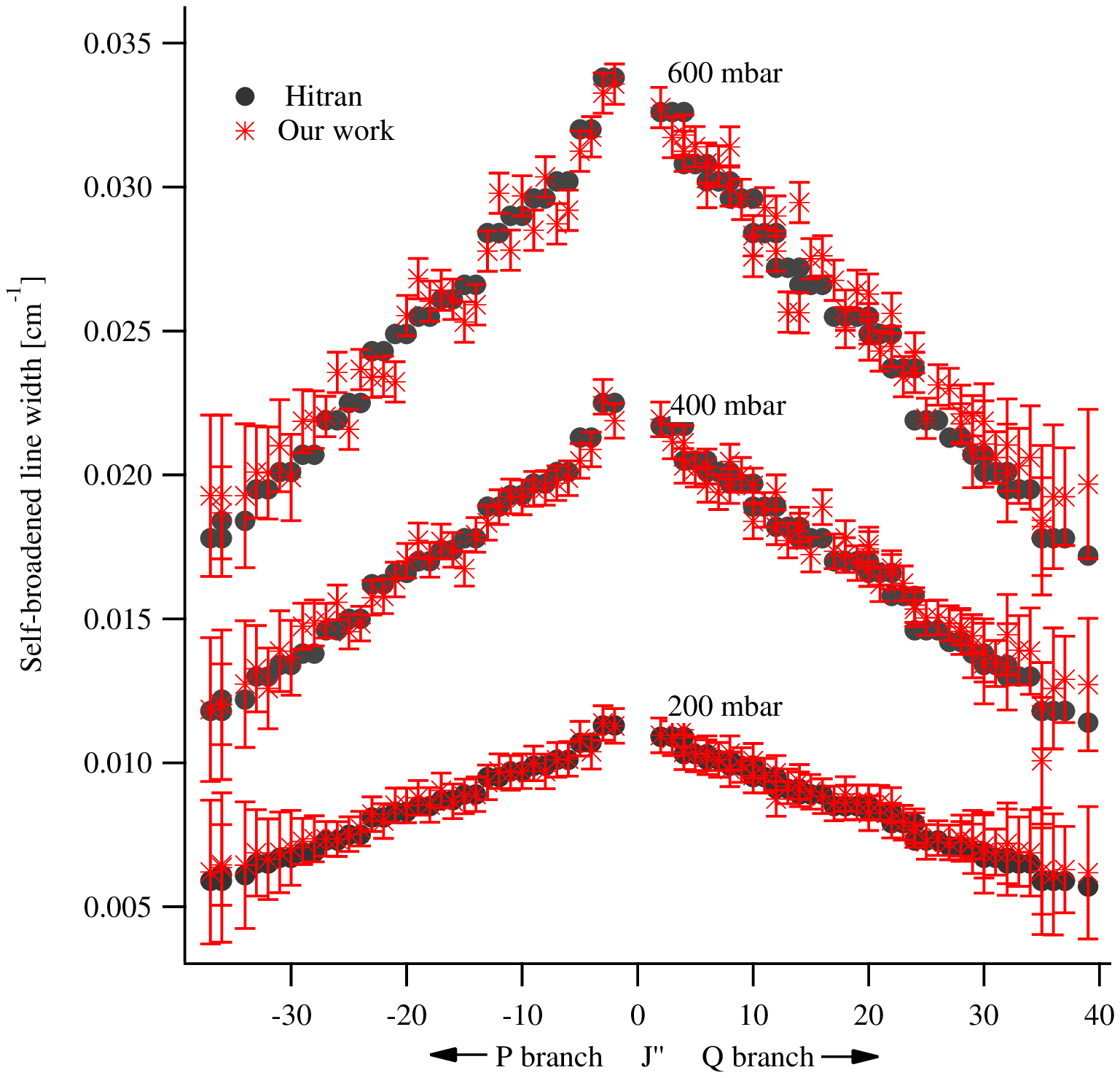}
  \caption{\label{o16PQselfbroadened} Width of $P$, $Q$ branch $^{16}$O$_2$.}
\end{minipage}%
\begin{minipage}{.65\textwidth}
  \centering
  \includegraphics[width=.8\linewidth]{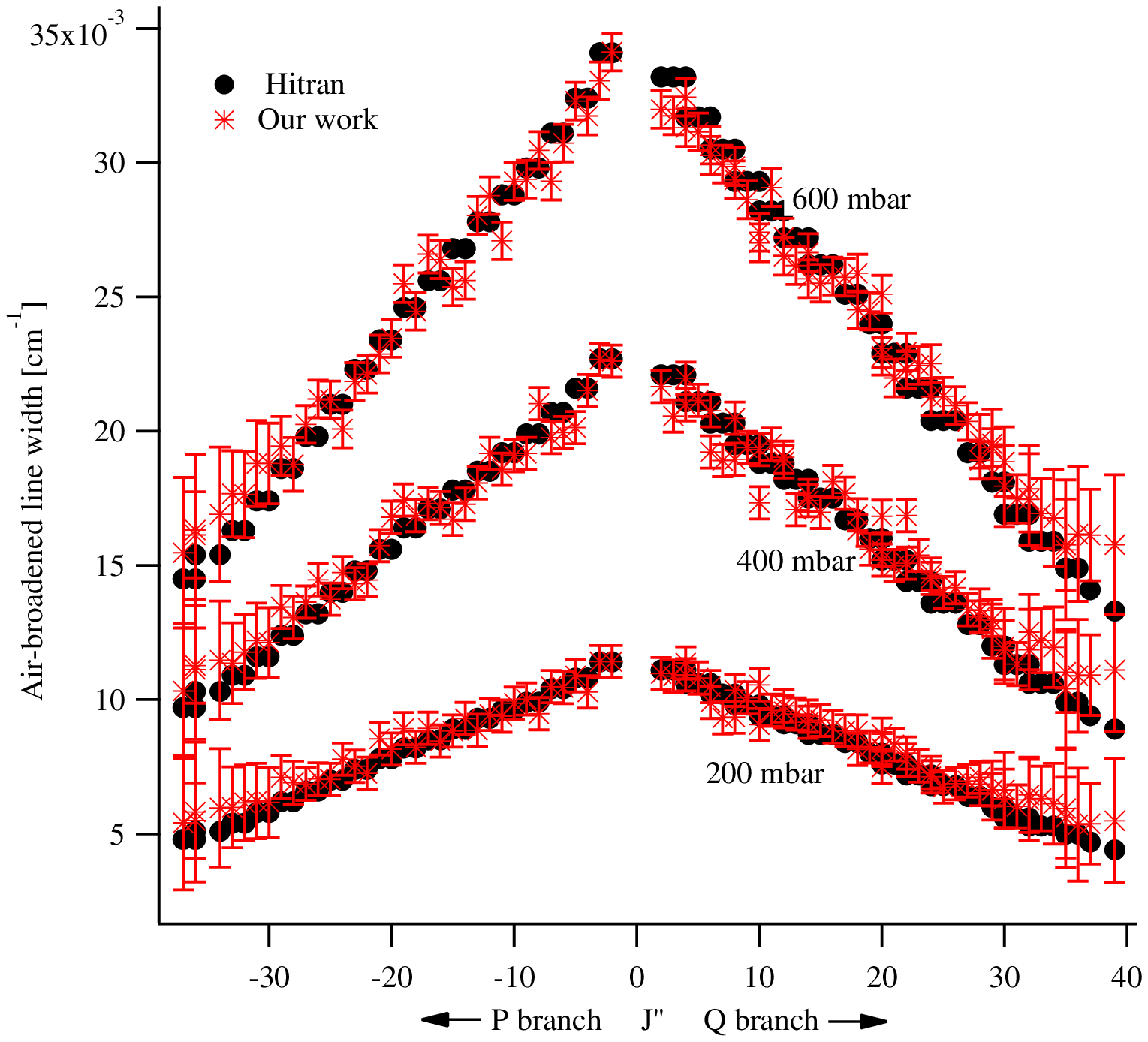}
  \caption{\label{o16PQairbroadened} Width of $P$, $Q$ branch $^{16}$O$_2$.}
\end{minipage}
\end{figure}

We have measured 14 transitions in the $O$ branch, 7 transition in
the $R$ branch, 36 transitions in the $P$ branch and 53 transition
in the $Q$ branch of $^{16}$O$_2$ in the
a${^1}{\triangle}_g{\leftarrow}$X $^3{\Sigma}^{-}_g$ $(0,0)$ band.
Our measured parameters and corresponding experimental uncertainty
of these transitions are given in Table~\ref{16O2Olineintensity},
Table~\ref{16O2Slineintensity}, Table~\ref{16O2Plineintensity} and
in Table~\ref{16O2Qlineintensity} respectively. The uncertainties
for some of the transitions are larger. Since many of the
transitions overlap and hence the fitting become more difficult.
In Fig.~\ref{o16O-Rbranch} and Fig.~\ref{o16P-Qbranch} we compared
our measured line intensities for these transitions with the
calculated values in the HITRAN database. Furthermore, we have
measured the self-broadened linewidth (measurement with ultra pure
oxygen sample) and the air-broadened linewidth (measurement with
dry air sample) for these transitions at three different pressure
i.e. 200 mbar, 400 mbar and 600 mbar which are shown in
Fig.~\ref{o16ORselfbroadened}, Fig.~\ref{o16ORairbroadened}
Fig.~\ref{o16PQselfbroadened} and in Fig.~\ref{o16PQairbroadened}
along with the calculated values in the HITRAN database. It is to
be noted that, the measured line positions, line intensities,
self-broadened and air-broadened Lorentzian half width are in good
agreement with the predictions in the HITRAN database.

\subsection{Lines of $^{18}$O$_2$}

\begin{figure}
\centering
\begin{minipage}{.65\textwidth}
  \centering
  \includegraphics[width=.9\linewidth]{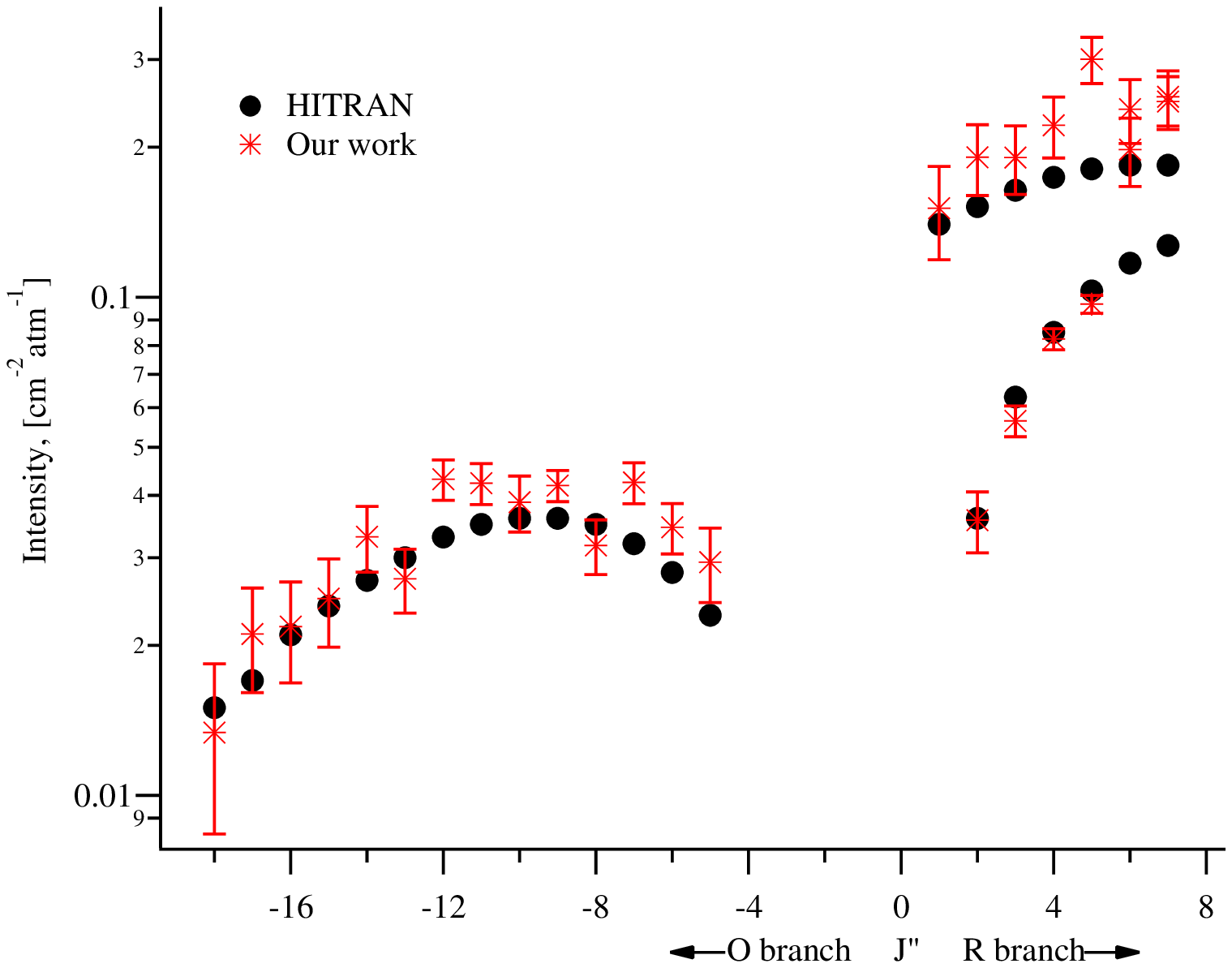}
  \caption{\label{o18O-Rbranch} Intensity of the $O$, $R$ branch of $^{18}$O$_2$.}
\end{minipage}%
\begin{minipage}{.65\textwidth}
  \centering
  \includegraphics[width=.8\linewidth]{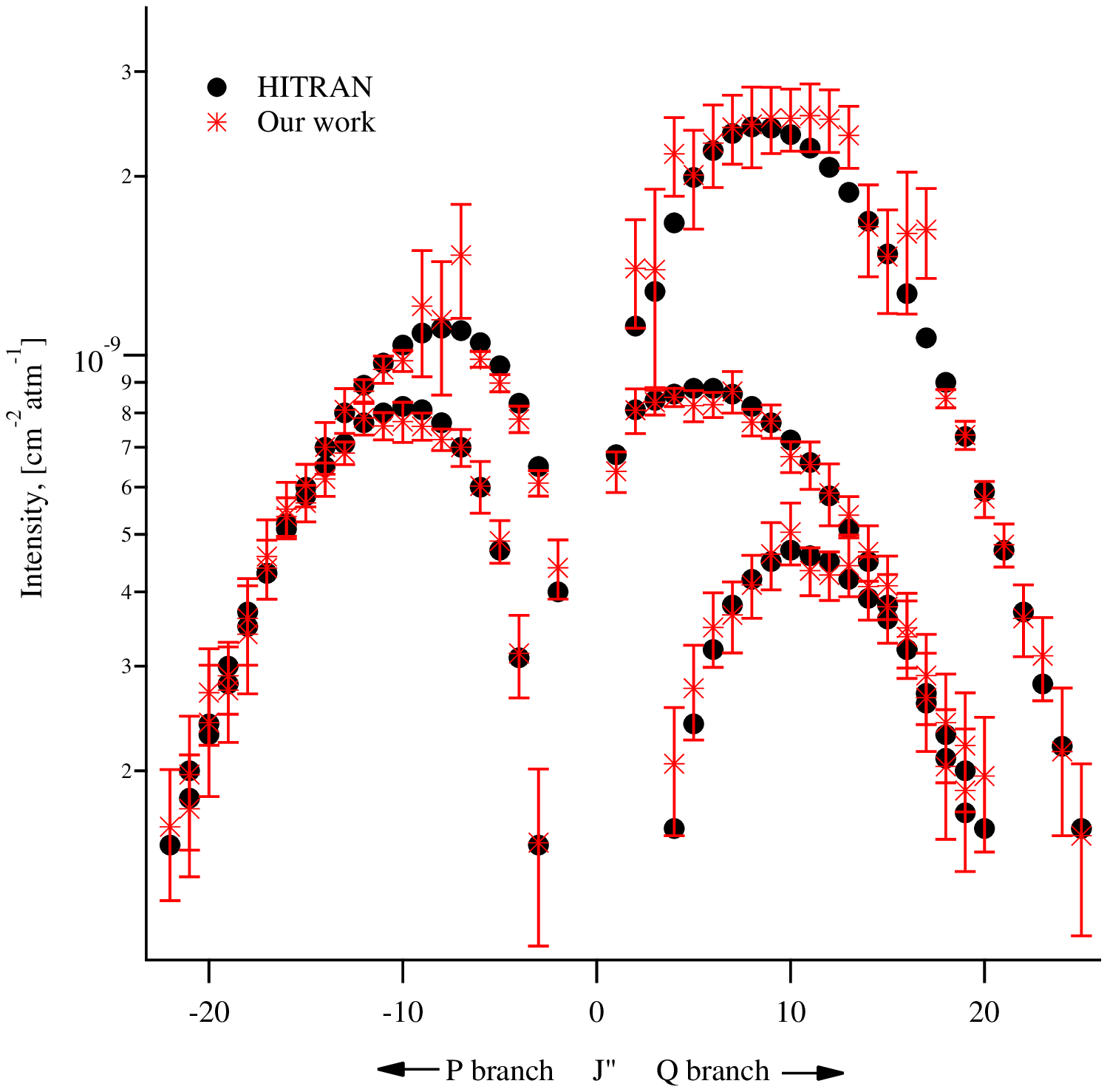}
  \caption{\label{o18P-Qbranch} Intensity in the $P$, $Q$ branch of $^{18}$O$_2$.}
\end{minipage}
\end{figure}

\begin{figure}
\centering
\begin{minipage}{.65\textwidth}
  \centering
  \includegraphics[width=.8\linewidth]{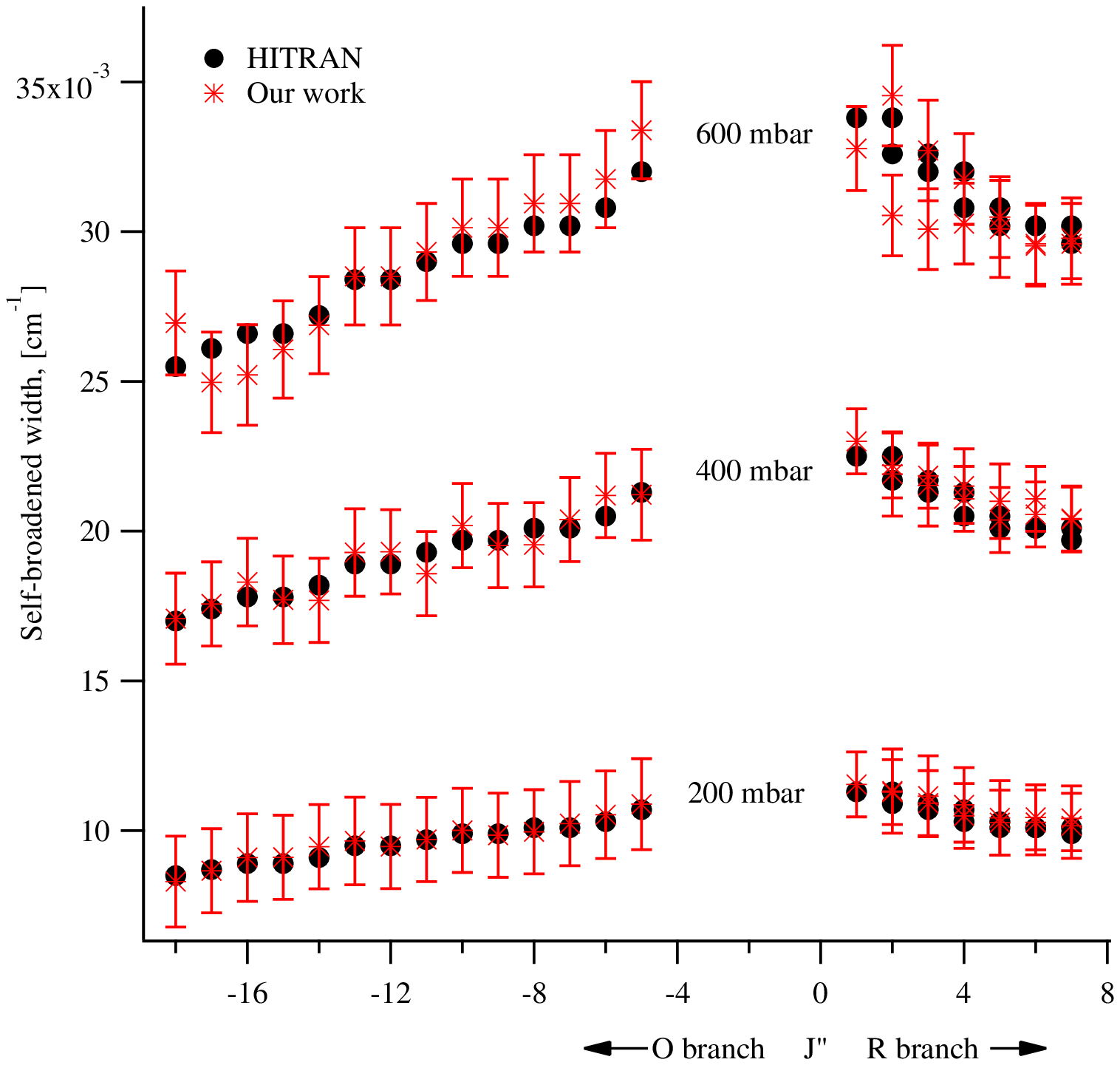}
  \caption{\label{o18ORselfbroadened} Width of $O$, $R$ branch $^{18}$O$_2$.}
\end{minipage}%
\begin{minipage}{.65\textwidth}
  \centering
  \includegraphics[width=.8\linewidth]{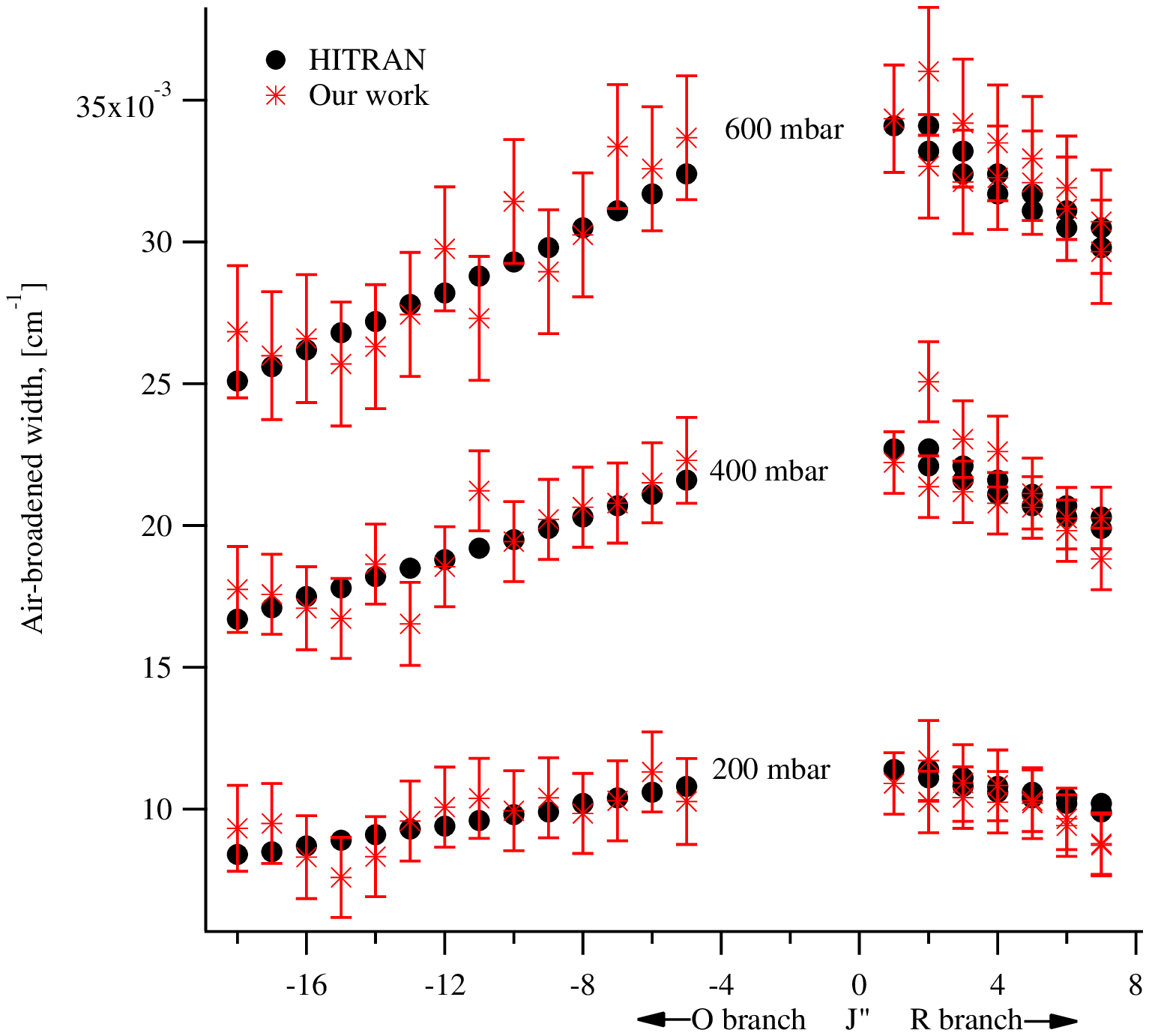}
  \caption{\label{o18ORairbroadened} Width of $O$, $R$ branch $^{18}$O$_2$.}
\end{minipage}
\end{figure}

\begin{figure}
\centering
\begin{minipage}{.65\textwidth}
  \centering
  \includegraphics[width=.8\linewidth]{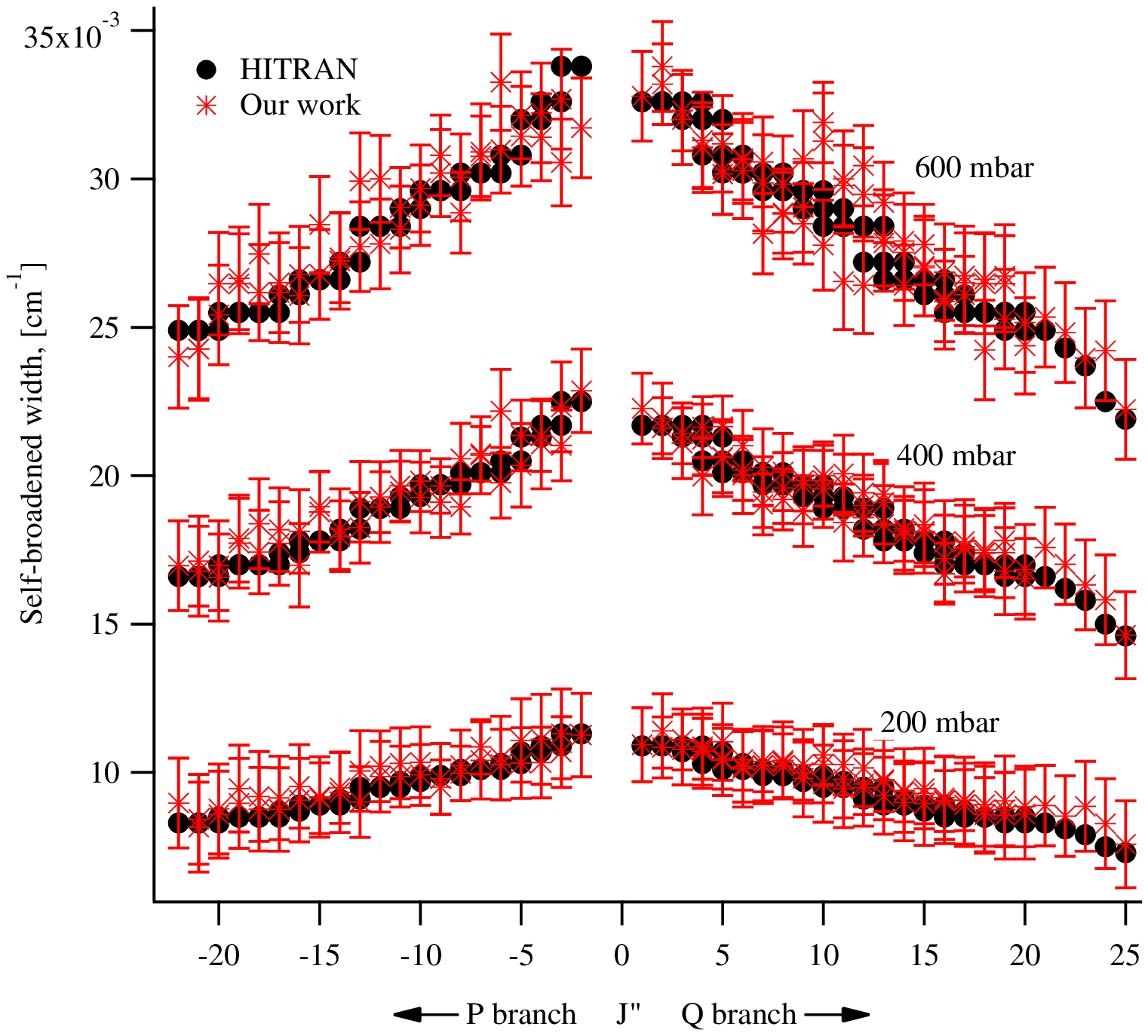}
  \caption{\label{o18PQselfbroadened} Width of $P$, $Q$ branch $^{18}$O$_2$.}
\end{minipage}%
\begin{minipage}{.65\textwidth}
  \centering
  \includegraphics[width=.8\linewidth]{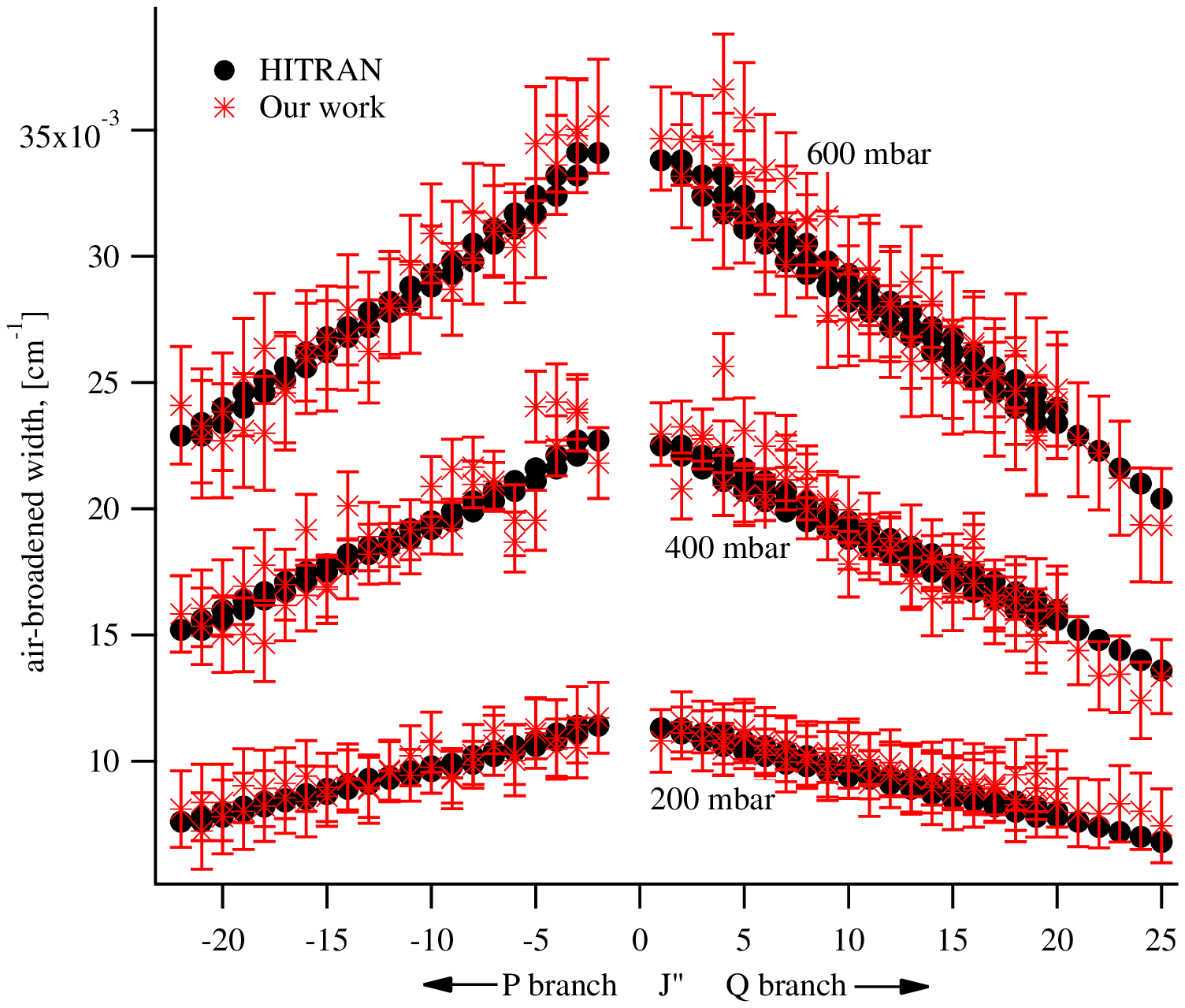}
  \caption{\label{o18PQairbroadened} Width of $P$, $Q$ branch $^{18}$O$_2$.}
\end{minipage}
\end{figure}

We have measured 14 transitions in the $O$ branch,13 transition in
the $R$ branch, 3 transition in the $S$ branch, 40 transitions in
the $P$ branch, and 60 transition in the $Q$ branch of
$^{18}$O$_2$ in the a${^1}{\triangle}_g{\leftarrow}$X
$^3{\Sigma}^{-}_g$ $(0,0)$ band. Our measured parameters and
corresponding experimental uncertainty of these transitions are
given in Table~\ref{18O2Olineintensity},
Table~\ref{18O2RSlineintensity}, Table~\ref{18O2Plineintensity}
and in Table~\ref{18O2Qlineintensity} respectively. In
Fig.~\ref{o18O-Rbranch}, Fig.~\ref{o18P-Qbranch} we compared our
measured line intensities for these transitions with the
calculated values in the HITRAN database. Furthermore, we have
measured the self-broadened linewidth (measurement with ultrapure
oxygen sample) and the air-broadened linewidth (measurement with
dry air sample) for these transitions at three different pressure
i.e. 200 mbar, 400 mbar and 600 mbar which are shown in
Fig.~\ref{o18ORselfbroadened}, Fig.~\ref{o18ORairbroadened}
Fig.~\ref{o18PQselfbroadened} and in Fig.~\ref{o18PQairbroadened}
along with the calculated values in the HITRAN database. It is to
be noted that, the measured line positions, line intensities,
self-broadened and air-broadened Lorentzian half width are in good
agreement with the predictions in the HITRAN database.

\subsection{Data fitting}
We used Matlab to perform a Voigt lineshape fit on each transition
to a theoretical NICE-OHMS lineshape function \cite{NLW04}. The
lineshape is accounts for the effect of the two levels of
frequency modulation.

The energy of a transition from the ground state $(N^{'},J^{'})$
to the excited state $(J)$ is given by the equation
\begin{equation}\label{fiteq}
\tilde{\nu}_{J\leftarrow N^{'},J^{'}} = \tilde{\nu}_{0\leftarrow
0,0} + B{_0}J(J+1)-D_0[J(J+1)]^{2}-F^{'}(N^{'},J^{'})
\end{equation}
where the lower state energy $F^{'}(N^{'}J^{'})$ is calculated
with the molecular constant in a recent publication \cite{OSI10}.
$\tilde{\nu}_{J\leftarrow N^{'},J^{'}}$ is the measured line
position, and $\tilde{\nu}_{0\leftarrow 0,0}$, $B{_0}$ and $D_0$
are the molecular constants of the excited state,
$a(\tilde{\nu}=0)$ which we measured.

\subsubsection{$^{16}$O$_{2}$ parameters}
We have measured 111 lines of the
a${^1}{\triangle}_g{\leftarrow}$X $^3{\Sigma}^{-}_g$ $(0,0)$ band
of $^{16}$O$_{2}$. To extract the parameters of each individual
transition we fit Eq.\ref{fiteq} to the line position of the
measured transition. The results of the fit are given in Table
\ref{016parcomp}, where the a${^1}{\triangle}_g$ state constants
compared with those from Olaga \cite{OSI10}, Amiot \cite{AMV81}
and Rothman \cite{ROT82}.

\subsubsection{$^{18}$O$_{2}$ parameters}
We have measured 130 lines of the
a${^1}{\triangle}_g{\leftarrow}$X $^3{\Sigma}^{-}_g$ $(0,0)$ band
of $^{18}$O$_{2}$. To extract the parameters of each individual
transition we fit Eq.\ref{fiteq} to the line position of the
measured transition. The results of the fit are given in Table
\ref{018parcomp}, where the a${^1}{\triangle}_g$ state constants
compared with those from Olaga \cite{OSI10}.

\section{Conclusion}
Our work characterized 241 ultraweak transitions in molecular
oxygen by measuring their line positions, self-broadened and
air-broadened linewidths and line intensities using NICE-OHMS. The
precise value of the parameters provide a path to improve the
existing HITRAN database. It is to be noted that the
a${^1}{\triangle}_g{\leftarrow}$X $^3{\Sigma}^{-}_g$ electric
quadrupole transitions of $^{16}$O$_{2}$ and $^{18}$O$_{2}$ are
important to atmospheric modelling. The NICE-OHMS measurement of
different a${^1}{\triangle}_g{\leftarrow}$X $^3{\Sigma}^{-}_g$
transitions in this work provide one of the best set of
spectroscopic parameters for the a${^1}{\triangle}_g$ state of
$^{16}$O$_{2}$ and $^{18}$O$_{2}$.

\section{Acknowledgements}
We are grateful to Tzahi Grunzweig for his help with labview for
automation of the experiment. This research is supported by the
New Zealand Foundation for Research, Science and Technology, Otago
University.

\pagebreak

\begin{table}
\caption{Measured transition parameters of the $O$ branch of
 $^{16}$O$_2$.}
\label{16O2Olineintensity}
\begin{tabular}{lcccc}
\hline\noalign{}
Transition & Position && Line intensity & Self-broadened half-width\\
$\Delta$ N N$^\prime$ $\Delta$ J J$^\prime$ & cm$^{-1}$ && 10$^{-8}$cm$^{-2}$atm$^{-1}$ & cm$^{-1}$ at 400 mbar\\
\noalign{}\hline\noalign{}
O31P30  &   7692.459(2) &&   0.139(18)   &   0.0136(14)  \\
O29P28  &   7706.069(2) &&   0.338(10)   &   0.0140(12)  \\
O27P26  &   7719.552(2) &&   0.722(14)   &   0.0144(9)   \\
O25P24  &   7732.894(2) &&   1.372(11)   &   0.0155(6)   \\
O23P22  &   7746.031(2) &&   3.062(10)   &   0.0163(6)   \\
O21P20  &   7759.092(2) &&   5.125(10)   &   0.0167(6)   \\
O19P18  &   7772.052(2) &&   7.369(10)   &   0.0167(6)   \\
O17P16  &   7784.807(2) &&   10.173(10)  &   0.0175(6)   \\
O15P14  &   7797.435(2) &&   13.147(10)  &   0.0179(6)   \\
O13P12  &   7809.917(2) &&   16.335(10)  &   0.0186(6)   \\
O11P10  &   7822.243(2) &&   18.934(10)  &   0.0194(6)   \\
O9P8    &   7834.445(2) &&   18.041(10)  &   0.0198(6)   \\
O7P6    &   7846.459(2) &&   14.563(10)  &   0.0202(6)   \\
O5P4    &   7858.330(2) &&   8.204(10)   &   0.0214(6)   \\
\noalign{\smallskip}\hline
\end{tabular}
\end{table}

\begin{table}
\caption{Measured transition parameters of the $R$ and $S$ branch
of $^{16}$O$_2$.} \label{16O2Slineintensity}
\begin{tabular}{lcccc}
\hline\noalign{}
Transition & Position && Line intensity & Self-broadened half width \\
$\Delta$ N N$^\prime$ $\Delta$ J J$^\prime$ & cm$^{-1}$ && 10$^{-8}$cm$^{-2}$atm$^{-1}$  &cm$^{-1}$  400 mbar \\
\noalign{}\hline\noalign{}
R1R1    &   7888.092(2) &&   71.688(10)  &   0.0222(6) \\
R1Q2    &   7889.930(2) &&   18.870(10)  &   0.0222(6) \\
R3R3    &   7893.558(2) &&   82.862(10)  &   0.0210(6) \\
R3Q4    &   7895.457(2) &&   41.828(10)  &   0.0210(6) \\
R5R5    &   7898.879(2) &&   88.711(10)  &   0.0198(6) \\
R5Q6    &   7900.841(2) &&   58.952(10)  &   0.0198(6) \\
R7R7    &   7904.023(2) &&   91.036(10)  &   0.0194(6) \\
\hline
S1R2    &   7898.470(2) &&   37.062(10)  &   0.0214(6)  \\
\noalign{\smallskip}\hline
\end{tabular}
\end{table}

\begin{center}
\begin{longtable}{cccc}
\caption{Measured transition parameters of the $P$ branch
of $^{16}$O$_2$.}\label{16O2Plineintensity}\\
\hline Transition & Position & Line intensity & Self-broadened
half-width\\
$\Delta$ N N$^\prime$ $\Delta$ J J$^\prime$ V& cm$^{-1}$ & 10$^{-8}$cm$^{-2}$atm$^{-1}$ & cm$^{-1}$ at 400 mbar\\
\hline
\endfirsthead
\multicolumn{4}{c}%
{\tablename\ \thetable\ -- \textit{Continued from previous page}} \\
\hline \hline Transition & Position & Line intensity & Self-broadened half-width\\
$\Delta$ N N$^\prime$ $\Delta$ J J$^\prime$ V& cm$^{-1}$ & 10$^{-8}$cm$^{-2}$atm$^{-1}$ & cm$^{-1}$ at 400 mbar\\
\hline
\endhead
\hline \multicolumn{4}{c}{\textit{Continued on next page}} \\
\endfoot
\hline
\endlastfoot
P37P37  &   7750.151(2) &   0.037(5)    &   0.0119(25)  \\
P37Q36  &   7751.843(2) &   0.032(5)    &   0.0120(16)  \\
P35P36  &   7758.676(2) &   0.209(17)   &   0.0120(14)  \\
P35Q34  &   7760.315(2) &   0.067(5)    &   0.0127(22)  \\
P33P33  &   7767.024(2) &   0.186(19)   &   0.0133(15)  \\
P33Q32  &   7768.760(2) &   0.148(22)   &   0.0126(14)  \\
P31P31  &   7775.219(2) &   0.318(17)   &   0.0139(14)  \\
P31Q30  &   7776.958(2) &   0.381(18)   &   0.0137(13)  \\
P29P29  &   7783.193(2) &   0.711(14)   &   0.0148(8)   \\
P29Q28  &   7785.02(2)  &   0.878(13)   &   0.0149(8)   \\
P27P27  &   7791.141(2) &   2.251(10)   &   0.0149(6)   \\
P27Q26  &   7792.931(2) &   1.622(10)   &   0.0156(6)   \\
P25P25  &   7798.861(2) &   3.629(10)   &   0.0146(6)   \\
P25Q24  &   7800.622(2) &   3.631(10)   &   0.0148(6)   \\
P23P23  &   7806.475(2) &   5.960(10)   &   0.0157(6)   \\
P23Q22  &   7808.213(2) &   6.657(10)   &   0.0158(6)   \\
P21P21  &   7813.861(2) &   10.867(10)  &   0.0164(6)   \\
P21Q20  &   7815.716(2) &   10.859(10)  &   0.0170(6)   \\
P19P19  &   7821.099(2) &   13.731(10)  &   0.0177(6)   \\
P19Q18  &   7822.939(2) &   16.290(10)  &   0.0171(6)   \\
P17P17  &   7828.202(2) &   20.672(10)  &   0.0177(6)   \\
P17Q16  &   7830.106(2) &   24.296(10)  &   0.0174(6)   \\
P15P15  &   7835.216(2) &   27.187(10)  &   0.0167(6)   \\
P15Q14  &   7837.104(2) &   33.127(10)  &   0.0179(6)   \\
P13P13  &   7841.963(2) &   34.795(10)  &   0.0183(6)   \\
P13Q12  &   7843.927(2) &   42.831(10)  &   0.0189(6)   \\
P11P11  &   7848.657(2) &   39.733(10)  &   0.0193(6)   \\
P11Q10  &   7850.604(2) &   49.796(10)  &   0.0192(6)   \\
P9P9    &   7855.182(2) &   40.231(10)  &   0.0195(6)   \\
P9Q8    &   7857.122(2) &   53.482(10)  &   0.0196(6)   \\
P7P7    &   7861.503(2) &   35.064(10)  &   0.0197(6)   \\
P7Q6    &   7863.463(2) &   52.423(10)  &   0.0199(6)   \\
P5P5    &   7867.711(2) &   23.739(10)  &   0.0205(6)   \\
P5Q4    &   7869.706(2) &   41.412(10)  &   0.0209(6)   \\
P3P3    &   7873.700(2) &   7.952(10)   &   0.0227(6)   \\
P3Q2    &   7875.785(2) &   19.828(10)  &   0.0219(6)   \\
\end{longtable}
\end{center}

\begin{center}
\begin{longtable}{cccc}
\caption{Measured transition parameters of the $Q$ branch
of $^{16}$O$_2$.}\label{16O2Qlineintensity}\\
\hline Transition & Position & Line intensity & Self-broadened
half-width\\
$\Delta$ N N$^\prime$ $\Delta$ J J$^\prime$ V& cm$^{-1}$ & 10$^{-8}$cm$^{-2}$atm$^{-1}$ & cm$^{-1}$ at 400 mbar\\
\hline
\endfirsthead
\multicolumn{4}{c}%
{\tablename\ \thetable\ -- \textit{Continued from previous page}} \\
\hline \hline Transition & Position & Line intensity & Self-broadened half-width\\
$\Delta$ N N$^\prime$ $\Delta$ J J$^\prime$ V& cm$^{-1}$ & 10$^{-8}$cm$^{-2}$atm$^{-1}$ & cm$^{-1}$ at 400 mbar\\
\hline
\endhead
\hline \multicolumn{4}{c}{\textit{Continued on next page}} \\
\endfoot
\hline
\endlastfoot
Q3R2    &   7884.298(2) &   40.095(10)  &   0.0219(6)   \\
Q3Q3    &   7882.217(2) &   64.156(10)  &   0.0212(6)   \\
Q3P4    &   7884.141(2) &   8.482(10)   &   0.0211(6)   \\
Q5R4    &   7883.870(2) &   43.470(10)  &   0.0203(6)   \\
Q5Q5    &   7881.834(2) &   99.431(10)  &   0.0202(6)   \\
Q5P6    &   7883.822(2) &   16.293(10)  &   0.0197(6)   \\
Q7R6    &   7883.285(2) &   43.599(10)  &   0.0203(6)   \\
Q7Q7    &   7881.326(2) &   116.910(10) &   0.0194(6)   \\
Q7P8    &   7883.318(2) &   20.976(10)  &   0.0205(6)   \\
Q9R8    &   7882.631(2) &   40.629(10)  &   0.0196(6)   \\
Q9Q9    &   7880.664(2) &   118.606(10) &   0.0200(6)   \\
Q9P10   &   7882.704(2) &   22.926(10)  &   0.0196(6)   \\
Q11R10  &   7881.713(2) &   35.100(10)  &   0.0184(6)   \\
Q11Q11  &   7879.818(2) &   107.894(10) &   0.0188(6)   \\
Q11P12  &   7881.892(2) &   22.012(10)  &   0.0194(6)   \\
Q13R12  &   7880.716(2) &   28.214(10)  &   0.0182(6)   \\
Q13Q13  &   7878.812(2) &   89.421(10)  &   0.0177(6)   \\
Q13P14  &   7880.908(2) &   19.084(10)  &   0.0181(6)   \\
Q15R14  &   7879.564(2) &   20.762(10)  &   0.0183(6)   \\
Q15Q15  &   7877.646(2) &   68.931(10)  &   0.0172(6)   \\
Q15P16  &   7879.798(2) &   15.592(10)  &   0.0189(6)   \\
Q17R16  &   7878.227(2) &   14.756(10)  &   0.0174(6)   \\
Q17Q17  &   7876.317(2) &   49.824(10)  &   0.0172(6)   \\
Q17P18  &   7878.490(2) &   11.131(10)  &   0.0178(6)   \\
Q19R18  &   7876.714(2) &   9.569(10)   &   0.0169(6)   \\
Q19Q19  &   7874.846(2) &   33.174(10)  &   0.0176(6)   \\
Q19P20  &   7877.008(2) &   8.197(10)   &   0.0174(6)   \\
Q21R20  &   7875.053(2) &   6.005(10)   &   0.0167(6)   \\
Q21Q21  &   7873.214(2) &   20.530(10)  &   0.0162(6)   \\
Q21P22  &   7875.399(2) &   5.245(10)   &   0.0168(6)   \\
Q23R22  &   7873.172(2) &   3.728(10)   &   0.0166(6)   \\
Q23Q23  &   7871.393(2) &   12.400(10)  &   0.0162(6)   \\
Q23P24  &   7873.571(2) &   3.341(10)   &   0.0153(6)   \\
Q25R24  &   7871.260(2) &   1.981(10)   &   0.0155(6)   \\
Q25Q25  &   7869.429(2) &   6.520(10)   &   0.0150(6)   \\
Q25P26  &   7871.593(2) &   2.112(10)   &   0.0151(6)   \\
Q27R26  &   7869.050(2) &   1.628(10)   &   0.0149(6)   \\
Q27Q27  &   7867.299(2) &   3.566(10)   &   0.0147(6)   \\
Q27P28  &   7869.517(2) &   0.783(13)   &   0.0146(8)   \\
Q29R28  &   7866.741(2) &   0.476(16)   &   0.0141(11)  \\
Q29Q29  &   7865.008(2) &   1.717(10)   &   0.0144(6)   \\
Q29P30  &   7867.212(2) &   0.401(16)   &   0.0139(11)  \\
Q31R30  &   7864.257(2) &   0.293(18)   &   0.0135(14)  \\
Q31Q31  &   7862.492(2) &   0.833(12)   &   0.0135(8)   \\
Q31P32  &   7864.728(2) &   0.186(19)   &   0.0145(14)  \\
Q33R32  &   7861.57(2)  &   0.165(19)   &   0.0133(15)  \\
Q33Q33  &   7859.856(2) &   0.342(17)   &   0.0139(12)  \\
Q33P34  &   7862.130(2) &   0.109(20)   &   0.0139(15)  \\
Q35R35  &   7858.739(2) &   0.072(6)    &   0.0101(22)  \\
Q35Q35  &   7857.004(2) &   0.157(20)   &   0.0120(15)  \\
Q35P36  &   7859.310(2) &   0.038(6)    &   0.0126(21)  \\
Q37Q37  &   7853.988(2) &   0.159(21)   &   0.0129(15)  \\
Q39Q39  &   7850.825(2) &   0.064(7)    &   0.0127(23)  \\
\end{longtable}
\end{center}

\begin{table}
\caption{Measured transition parameters of the $O$ branch of
 $^{18}$O$_2$.}
\label{18O2Olineintensity}
\begin{tabular}{lcccc}
\hline\noalign{}
Transition & Position && Line intensity & Self-broadened half-width\\
$\Delta$ N N$^\prime$ $\Delta$ J J$^\prime$ V& cm$^{-1}$ && 10$^{-8}$cm$^{-2}$atm$^{-1}$ & cm$^{-1}$ at 400 mbar\\
\noalign{}\hline\noalign{}
O19P18  &   7779.627(2) &&   0.013(5)    &   0.0171(15)  \\
O18P17  &   7785.657(2) &&   0.021(5)    &   0.0176(14)  \\
O17P16  &   7791.669(2) &&   0.022(5)    &   0.0183(15)  \\
O16P15  &   7797.647(2) &&   0.025(5)    &   0.0177(15)  \\
O15P14  &   7803.528(2) &&   0.033(5)    &   0.0177(14)  \\
O14P13  &   7809.441(2) &&   0.027(4)    &   0.0193(15)  \\
O13P12  &   7815.363(2) &&   0.043(4)    &   0.0193(14)  \\
O12P11  &   7821.190(2) &&   0.042(4)    &   0.0186(14)  \\
O11P10  &   7827.009(2) &&   0.039(5)    &   0.0202(14)  \\
O10P9   &   7832.777(2) &&   0.042(3)    &   0.0195(14)  \\
O9P8    &   7838.509(2) &&   0.032(4)    &   0.0195(14)  \\
O8P7    &   7844.231(2) &&   0.042(4)    &   0.0204(14)  \\
O7P6    &   7849.877(2) &&   0.035(4)    &   0.0212(14)  \\
O6P5    &   7855.489(2) &&   0.029(5)    &   0.0212(14)  \\
\noalign{\smallskip}\hline
\end{tabular}
\end{table}

\begin{table}
\caption{Measured transition parameters of the $R$ and $S$ branch
of $^{18}$O$_2$.} \label{18O2RSlineintensity}
\begin{tabular}{lcccc}
\hline\noalign{}
Transition & Position && Line intensity & Self-broadened half-width\\
$\Delta$ N N$^\prime$ $\Delta$ J J$^\prime$ V& cm$^{-1}$ && 10$^{-8}$cm$^{-2}$atm$^{-1}$ & cm$^{-1}$ at 400 mbar\\
\noalign{}\hline\noalign{}
R1R1    &   7889.051(2) &&   0.151(32)   &   0.0230(11)  \\
R1Q2    &   7890.971(2) &&   0.036(5)    &   0.0219(14)  \\
R2R2    &   7891.679(2) &&   0.191(31)   &   0.0222(11)  \\
R2Q3    &   7893.593(2) &&   0.056(4)    &   0.0215(14)  \\
R3R3    &   7894.221(2) &&   0.191(30)   &   0.0219(11)  \\
R3Q4    &   7896.199(2) &&   0.082(4)    &   0.0215(12)  \\
R4R4    &   7896.781(2) &&   0.221(31)   &   0.0211(11)  \\
R4Q5    &   7898.773(2) &&   0.097(4)    &   0.0210(12)  \\
R5R5    &   7899.278(2) &&   0.300(32)   &   0.0204(11)  \\
R5Q6    &   7901.264(2) &&   0.238(35)   &   0.0211(11)  \\
R6R6    &   7901.729(2) &&   0.198(31)   &   0.0206(11)  \\
R6Q7    &   7903.717(2) &&   0.253(32)   &   0.0204(11)  \\
R7R7    &   7904.157(2) &&   0.247(30)   &   0.0204(11)  \\
\hline
S0R1    &   7893.632(2) &&   0.059(4)    &   0.0225(12)  \\
S1R2    &   7899.001(2) &&   0.070(4)    &   0.0219(12)  \\
S2R3    &   7904.325(2) &&   0.076(4)    &   0.0214(12)  \\
\noalign{\smallskip}\hline
\end{tabular}
\end{table}

\begin{center}
\begin{longtable}{lccc}
\caption{Measured transition parameters of the $P$ branch
of $^{18}$O$_2$.}\label{18O2Plineintensity}\\
\hline Transition & Position & Line intensity & Self-broadened
half-width\\
$\Delta$ N N$^\prime$ $\Delta$ J J$^\prime$ V& cm$^{-1}$ & 10$^{-8}$cm$^{-2}$atm$^{-1}$ & cm$^{-1}$ at 400 mbar\\
\hline
\endfirsthead
\multicolumn{4}{c}%
{\tablename\ \thetable\ -- \textit{Continued from previous page}} \\
\hline \hline Transition & Position & Line intensity & Self-broadened half-width\\
$\Delta$ N N$^\prime$ $\Delta$ J J$^\prime$ V& cm$^{-1}$ & 10$^{-8}$cm$^{-2}$atm$^{-1}$ & cm$^{-1}$ at 400 mbar\\
\hline
\endhead
\hline \multicolumn{4}{c}{\textit{Continued on next page}} \\
\endfoot
\hline
\endlastfoot
P22P22  &   7815.521(2) &   0.016(4)    &   0.0170(15)  \\
P22Q21  &   7817.345(2) &   0.017(4)    &   0.0168(15)  \\
P21P21  &   7818.981(2) &   0.020(5)    &   0.0171(15)  \\
P21Q20  &   7820.831(2) &   0.024(6)    &   0.0170(15)  \\
P20P20  &   7822.459(2) &   0.027(5)    &   0.0166(15)  \\
P20Q19  &   7824.258(2) &   0.027(5)    &   0.0179(15)  \\
P19P19  &   7825.796(2) &   0.029(4)    &   0.0177(15)  \\
P19Q18  &   7827.690(2) &   0.034(7)    &   0.0184(15)  \\
P18P18  &   7829.210(2) &   0.036(6)    &   0.0174(14)  \\
P18Q17  &   7831.084(2) &   0.044(5)    &   0.0182(14)  \\
P17P17  &   7832.571(2) &   0.046(7)    &   0.0177(14)  \\
P17Q16  &   7834.393(2) &   0.055(6)    &   0.0181(14)  \\
P16P16  &   7835.876(2) &   0.054(4)    &   0.0170(14)  \\
P16Q15  &   7837.756(2) &   0.056(4)    &   0.0190(12)  \\
P15P15  &   7839.164(2) &   0.061(5)    &   0.0188(14)  \\
P15Q14  &   7841.035(2) &   0.070(7)    &   0.0180(12)  \\
P14P14  &   7842.343(2) &   0.062(4)    &   0.0182(14)  \\
P14Q13  &   7844.284(2) &   0.081(7)    &   0.0182(12)  \\
P13P13  &   7845.528(2) &   0.068(3)    &   0.0191(14)  \\
P13Q12  &   7847.487(2) &   0.087(4)    &   0.0193(12)  \\
P12P12  &   7848.723(2) &   0.078(5)    &   0.0189(12)  \\
P12Q11  &   7850.656(2) &   0.095(5)    &   0.0195(10)  \\
P11P11  &   7851.817(2) &   0.076(4)    &   0.0197(12)  \\
P11Q10  &   7853.783(2) &   0.098(4)    &   0.0198(10)  \\
P10P10  &   7854.912(2) &   0.077(6)    &   0.0193(12)  \\
P10Q9   &   7856.884(2) &   0.121(29)   &   0.0197(9)   \\
P9P9    &   7858.003(2) &   0.076(4)    &   0.0191(12)  \\
P9Q8    &   7859.914(2) &   0.115(29)   &   0.0190(9)   \\
P8P8    &   7861.020(2) &   0.072(3)    &   0.0206(12)  \\
P8Q7    &   7862.962(2) &   0.147(32)   &   0.0207(9)   \\
P7P7    &   7863.980(2) &   0.070(5)    &   0.0207(13)  \\
P7Q6    &   7865.945(2) &   0.098(3)    &   0.0198(12)  \\
P6P6    &   7866.943(2) &   0.060(6)    &   0.0222(14)  \\
P6Q5    &   7868.863(2) &   0.090(3)    &   0.0214(12)  \\
P5P5    &   7869.786(2) &   0.049(4)    &   0.0203(14)  \\
P5Q4    &   7871.769(2) &   0.078(4)    &   0.0213(12)  \\
P4P4    &   7872.687(2) &   0.032(5)    &   0.0211(15)  \\
P4Q3    &   7874.676(2) &   0.061(3)    &   0.0210(12)  \\
P3P3    &   7875.457(2) &   0.015(5)    &   0.0223(15)  \\
P3Q2    &   7877.569(2) &   0.044(5)    &   0.0229(14)  \\
\end{longtable}
\end{center}

\begin{center}
\begin{longtable}{lccc}
\caption{Measured transition parameters of the $Q$ branch
of $^{18}$O$_2$.}\label{18O2Qlineintensity}\\
\hline Transition & Position & Line intensity & Self-broadened
half-width\\
$\Delta$ N N$^\prime$ $\Delta$ J J$^\prime$ V& cm$^{-1}$ & 10$^{-8}$cm$^{-2}$atm$^{-1}$ & cm$^{-1}$ at 400 mbar\\
\hline
\endfirsthead
\multicolumn{4}{c}%
{\tablename\ \thetable\ -- \textit{Continued from previous page}} \\
\hline \hline Transition & Position & Line intensity & Self-broadened half-width\\
$\Delta$ N N$^\prime$ $\Delta$ J J$^\prime$ V& cm$^{-1}$ & 10$^{-8}$cm$^{-2}$atm$^{-1}$ & cm$^{-1}$ at 400 mbar\\
\hline
\endhead
\hline \multicolumn{4}{c}{\textit{Continued on next page}} \\
\endfoot
\hline
\endlastfoot
Q2R1    &   7885.814(2) &   0.064(5)    &   0.0223(12)  \\
Q2Q2    &   7883.661(2) &   0.140(29)   &   0.0216(10)  \\
Q3R2    &   7885.623(2) &   0.081(7)    &   0.0219(12)  \\
Q3Q3    &   7883.539(2) &   0.139(51)   &   0.0214(10)  \\
Q3P4    &   7885.485(2) &   0.021(5)    &   0.0200(13)  \\
Q4R3    &   7885.414(2) &   0.083(4)    &   0.0211(12)  \\
Q4Q4    &   7883.378(2) &   0.218(33)   &   0.0216(10)  \\
Q4P5    &   7885.322(2) &   0.028(5)    &   0.0206(13)  \\
Q5R4    &   7885.216(2) &   0.085(3)    &   0.0212(12)  \\
Q5Q5    &   7883.168(2) &   0.201(38)   &   0.0217(10)  \\
Q5P6    &   7885.191(2) &   0.035(5)    &   0.0200(13)  \\
Q6R5    &   7884.96(2)  &   0.082(5)    &   0.0206(12)  \\
Q6Q6    &   7882.980(2) &   0.227(36)   &   0.0202(10)  \\
Q6P7    &   7884.975(2) &   0.037(5)    &   0.0203(13)  \\
Q7R6    &   7884.664(2) &   0.083(4)    &   0.0210(12)  \\
Q7Q7    &   7882.689(2) &   0.242(32)   &   0.0190(10)  \\
Q7P8    &   7884.695(2) &   0.041(5)    &   0.0201(13)  \\
Q8R7    &   7884.331(2) &   0.087(7)    &   0.0195(12)  \\
Q8Q8    &   7882.443(2) &   0.245(38)   &   0.0192(10)  \\
Q8P9    &   7884.394(2) &   0.046(6)    &   0.0197(13)  \\
Q9R8    &   7884.052(2) &   0.077(4)    &   0.0196(12)  \\
Q9Q9    &   7882.072(2) &   0.250(32)   &   0.0198(10)  \\
Q9P10   &   7884.124(2) &   0.050(6)    &   0.0198(13)  \\
Q10R9   &   7883.598(2) &   0.078(5)    &   0.0188(12)  \\
Q10Q10  &   7881.699(2) &   0.250(30)   &   0.0200(10)  \\
Q10P11  &   7883.744(2) &   0.043(4)    &   0.0201(13)  \\
Q11R10  &   7883.19(2)  &   0.068(4)    &   0.0196(13)  \\
Q11Q11  &   7881.273(2) &   0.253(33)   &   0.0197(10)  \\
Q11P12  &   7883.342(2) &   0.043(4)    &   0.0194(13)  \\
Q12R11  &   7882.734(2) &   0.065(6)    &   0.0184(13)  \\
Q12Q12  &   7880.819(2) &   0.249(30)   &   0.0190(10)  \\
Q12P13  &   7882.927(2) &   0.044(5)    &   0.0191(14)  \\
Q13R12  &   7882.263(2) &   0.059(7)    &   0.0186(13)  \\
Q13Q13  &   7880.326(2) &   0.234(28)   &   0.0194(10)  \\
Q13P14  &   7882.433(2) &   0.041(5)    &   0.0182(14)  \\
Q14R13  &   7881.726(2) &   0.054(4)    &   0.0184(13)  \\
Q14Q14  &   7879.845(2) &   0.165(29)   &   0.0181(10)  \\
Q14P15  &   7881.898(2) &   0.038(5)    &   0.0183(14)  \\
Q15R14  &   7881.138(2) &   0.047(5)    &   0.0180(13)  \\
Q15Q15  &   7879.247(2) &   0.147(29)   &   0.0181(10)  \\
Q15P16  &   7881.361(2) &   0.034(5)    &   0.0178(14)  \\
Q16R15  &   7880.536(2) &   0.041(5)    &   0.0181(14)  \\
Q16Q16  &   7878.644(2) &   0.160(43)   &   0.0167(10)  \\
Q16P17  &   7880.767(2) &   0.027(5)    &   0.0178(14)  \\
Q17R16  &   7879.864(2) &   0.035(5)    &   0.0172(15)  \\
Q17Q17  &   7878.009(2) &   0.163(28)   &   0.0176(10)  \\
Q17P18  &   7880.122(2) &   0.024(5)    &   0.0175(14)  \\
Q18R17  &   7879.166(2) &   0.029(5)    &   0.0175(15)  \\
Q18Q18  &   7877.326(2) &   0.085(3)    &   0.0174(12)  \\
Q18P19  &   7879.414(2) &   0.022(5)    &   0.0174(15)  \\
Q19R18  &   7878.482(2) &   0.020(5)    &   0.0174(15)  \\
Q19Q19  &   7876.581(2) &   0.073(4)    &   0.0178(12)  \\
Q19P20  &   7878.708(2) &   0.020(5)    &   0.0168(15)  \\
\end{longtable}
\end{center}

\pagebreak

\begin{table}
\caption{Comparison of molecular constants in cm$^{-1}$ of the
$\nu$ = 0 levels of the a${^1}{\triangle}_g$ state of
$^{16}$O$_{2}$.} \label{016parcomp}
\begin{tabular}{lllll}
\hline\noalign{}
Constants & This work & Olaga \cite{OSI10} & Amiot \cite{AMV81} & Rothman \cite{ROT82} \\
\noalign{}\hline\noalign{}
$\tilde{\nu}_{0\leftarrow 0,0}$  & 7883.756662(142)  & 7883.756645(113)   &   7883.76179(28) &   7882.4288(3) \\
$B{_0}$                          & 1.417832166 (66)  & 1.417839039 (38)   &   1.41784020(82) &   1.4178442(19) \\
$D{_0}\times$ 10$^{-6}$          & 5.102159(351)     & 5.102256(243)      &   5.1074(24)     &   5.11144(139)  \\
\noalign{\smallskip}\hline
\end{tabular}
\end{table}

\begin{table}
\caption{Comparison of molecular constants in cm$^{-1}$ of the
$\nu$ = 0 levels of the a${^1}{\triangle}_g$ state of
$^{18}$O$_{2}$.} \label{018parcomp}
\begin{tabular}{lll}
\hline\noalign{}
Constants & This work & Olaga \cite{OSI10} \\
\noalign{}\hline\noalign{}
$\tilde{\nu}_{0\leftarrow 0,0}$  & 7886.512585(200)  & 7886.409277(117)   \\
$B{_0}$                          & 1.260406188(110)  & 1.260409499 (56)   \\
$D{_0}\times$ 10$^{-6}$          & 4.011059(850)     & 4.029678(664)      \\
\noalign{\smallskip}\hline
\end{tabular}
\end{table}

\bibliographystyle{elsarticle-num}
\bibliography{<your-bib-database>}



\end{document}